**Comments on "DIRECT-Net: A unified mutual-domain material decomposition network for quantitative dual-energy CT imaging"**

Xiaochuan Pan[1,2] and Emil Sidky[1]

[1]Department of Radiology and [2]Department of Radiation & Cellular Oncology, The University of Chicago, Chicago, USA

Quantitative image reconstruction in dual-energy computed tomography (CT) remains a topic of active research. We read with interest "DIRECT-Net: A unified mutual-domain material decomposition network for quantitative dual-energy CT imaging," which, referred to as the paper hereinafter, appears in the 2022 February Issue of *Med Phys*.[1] In the paper the authors propose a deep-learning (DL) method, referred to as the *Direct-Net* method, to address the problem of quantitative image reconstruction <u>directly</u> from data in full-scan dual-energy CT (DECT). We would like to comment on the study and conclusion in the paper. (Words and numbers in *italic* below are quoted from the paper.)

1. **Evidence in support of the conclusion**

There exists an extensive body of research works in the literature on theoretical development and practical application of methods that can address well the problem. The paper concludes that the "*DIRECT-Net has promising benefits in improving the DECT image quality*." As the conclusion stands, evidence is needed for demonstrating the "*benefits*" of the *Direct-Net* method proposed for improving over, i.e., performing better than, relevant existing methods addressing the same problem. As commented below, however, no meaningful evidence can be identified in the paper supporting the conclusion.

2. **Results of comparison methods**

The level of relative errors in concentration estimations for the calcium chloride and iodine solution phantoms in the paper, as reported in *Tables 2 & 3*, appears notably inconsistent with that of relative errors reported in the literature for the methods compared. Much of the relative errors obtained in the paper for the comparison methods (references cited), i.e., *ID-EP (Ref. [7])*,[2] *Direct-JSI (Ref. [14])*,[3] *Butterfly-Net (Ref. [20])*,[4] and *Clark-Net (Ref. [23])*,[5] are over 20% in *Table 2* and over 10% in *Table 3*, with maximum relative errors > 90%. In contrast, in Table II of *Ref. [7]*,[2] an average relative error of 4.2% with a max of 12.83% has been reported; in Table II of *Ref. [14]*,[3] relative errors can be computed as <1% for bone and tissue regions; in Table I of *Ref. [20]*,[4] relative errors can be estimated to be <1%; and in *Ref. [23]*,[5] the relative error has been reported as about 3%. While the study conditions in *Refs. [7]*,[2] *[14]*,[3] *[20]*,[4] and *[23]*[5] may not be identical to those in the paper, the difference in estimation errors between the Direct-Net method and the comparison methods appears perplexingly large that it calls for proper explanation.*

Accurate concentration estimates depend primarily on accurate physics modeling of the measurement, and estimation error can occur when the modeled physics such as spectra is not the same as the physics in simulation and/or experiment DECT. It is unclear if there is any mismatch between the physics in the data and the modeled physics in the methods compared.

For the non-DL methods considered in the paper, results are obtained with an *image-domain decomposition (IDD)* method, referred to as the *ID-EP* method *(Ref. [7])*,[2] and *a direct*

*iterative decomposition* method, referred to as the *Direct-JSI* method *(Ref. [14])*.[3] The concentration estimations of the Direct-JSI method are much worse than those of the ID-EP method in the paper. The former appears to use the proper physics modeling, while the latter makes a linear approximation of the logarithm of *equation (1)* in the paper; and one would thus expect that the performance of the former is at least as good as that of the latter. It is unclear as to why the result of the Direct-JSI method appears considerably worse than that of the ID-EP method.

3. **Study conditions**

It can be observed that the low- and high-kVp FBP reconstructions in *Figs. 6(a), 7(a),* and *8(a)* in the paper display little visible beam-hardening artifacts (possibly due to the small cross-section size and material composition of the phantoms and knuckle specimen used.) As such, accurate concentration estimations of calcium and iodine can be expected, respectively, by use of the standard IDD method from the FBP reconstructions (with a simple calibration if necessary), the well-established data-domain decomposition (DDD) methods, and the recent one-step algorithms. In other words, the study with insignificant beam-hardening effect demonstrates neither that the Direct-Net method proposed can accurately reconstruct images in full-scan DECT with a realistic level of beam-hardening effect nor that it performs better than do the relevant, existing methods for full-scan DECT. It is indeed unclear why concentration estimations with significant differences were obtained by use of the methods compared for this case only with a weak beam-hardening effect.

The Direct-Net method's performance evaluated under the condition of a realistic level of beam-hardening effect, clearly more prominent than that considered in the paper, would be necessary to support the paper's conclusion.

4. **Comparison with relevant, state-of-the-art methods**

Numerous methods have been reported in the literature for quantitative image reconstruction and accurate concentration estimations in full-scan DECT. To support the paper's conclusion, the Direct-Net method logically needs to be compared with the relevant, established methods, including the three types of the IDD methods, the DDD methods, and especially the recent one-step algorithms because both Direct-Net method and recent one-step algorithms reconstruct images <u>directly</u> from data in full-scan DECT. The statement in the paper that "*Comparisons are performed with different DECT decomposition algorithms*" is not supported by sufficient evidence because the Direct-Net method was not compared appropriately with the relevant, existing algorithms or methods in each of the three types.

The standard IDD method, the most widely used in clinical and other applications, directly inverts a 2X2 matrix of equation (1) in *Ref. [7]*[2] for obtaining basis images from low- and high-kVp FBP reconstructions. It subsequently estimates material concentrations from the basis images obtained. No result can be found in the paper comparing the Direct-Net and standard IDD methods.

The DDD methods are used widely in academic research on, and industrial products of, DECT to minimize effectively beam-hardening artifacts, thus yielding accurate concentration estimation. No result can be found in the paper comparing the Direct-Net and DDD methods.

The recent one-step algorithms have been demonstrated to yield accurate estimations of material concentrations in not only full-scan, but also partial-scan, DECT. While both the Direct-Net method and one-step algorithms reconstruct basis images directly from data in full-scan DECT, no results are presented in the paper comparing the Direct-Net method with the recent one-step algorithms, and appropriate citations of relevant, recent one-step algorithms appear missing in the paper.

5. **Reconstruction parameters**

It is unclear that the performance rankings of the methods compared in the paper are conclusive. Because a method such as those considered in the paper involves parameters (please also see Comment 7c below,) its reconstructed image, concentration estimation, and any goodness metrics are thus functions in a multi-dimensional space formed by the parameters involved. The results presented in the paper were obtained only at a single point in each of the respective multi-dimensional parameter spaces of the methods considered. Therefore, the conclusion about the relative rankings of their performance obtained from these single-parameter-point results is unlikely to be conclusive. This is because, when different points are chosen in these parameter spaces, the results and performance rankings of the methods are likely to vary, possibly leading to observations and conclusion contrary to those made in the paper.

6. **Solving the DECT inverse problem**

In DECT or spectral CT, the relationship between basis functions and model data are characterized often by imaging model *equation (1)* in the paper. In terms of basic research on reconstruction algorithms, quantitative image reconstruction is referred to often as accurate inversion of imaging model *equation (1)*. Numerical evidence has been reported in the literature that the recent one-step algorithms quantitatively reconstruct images by accurately inverting *equation (1)* directly from model data in full-scan and partial-scan DECT. No evidence can be found in the paper showing the "*benefits*" of the Direct-Net method for "*improving*" over the recent one-step algorithms in accurate image reconstruction in terms of accurate inversion of *equation (1)* from full-scan or partial-scan data in DECT.

7. **The paper's comments on one-step algorithms**

It was stated in the paper that the one-step algorithms "*may have strong spectra dependence and parameter selection issues, and they may also require lon time.*" The statement is unsubstantiated, as discussed below:

7a. The development of recent one-step algorithms for DECT was motivated by the design and enabling of novel DECT scan configurations. The research presented in the paper is for image reconstruction in DECT with a standard, full-scan configuration that is well-studied, and its relevance and significance thus diminish in terms of researching on novel algorithms for image reconstruction in full-scan DECT.**

7b. It is unclear what "*strong spectra dependence*" precisely is referred to. If the paper suggests that the robustness of the recent one-step algorithms is highly sensitive to the choice of spectra, no evidence can be identified in the paper to support the suggestion. In contrary, works in the literature reveal that the recent one-step algorithms are not more sensitive to the choice of spectra than any other methods in terms of accurate image reconstruction in full-scan (and partial scan) DECT.

7c. Any methods, including the Direct-Net method proposed and non-DL methods, will involve parameters and their value selections. For example, in the Direct-Net method proposed,[1] the design of the subnetworks and the DT module involves a slew of parameters listed in *Table 1* and the specific forms indicated in *equations (3)* and *(4)*; equally importantly, the design and implementation of the loss function, training schemes, and training/target image curation also involve multiple parameters and choices; there are additional parameter selections as discussed in the paragraph below *equation (7)*; and even the Direct-Net's architecture design itself constitutes a parameter of choice, as the "*room to improve the DIRECT-Net*" architecture was discussed in *Sec. 5* of the paper. In fact, given data, image reconstruction of any methods involves parameters, including the choices of image representations such as voxels and blobs and their sizes, imaging model forms, tube spectra and detector response, objective functions, algorithms, and network parameters. Therefore, like any reconstruction methods, the Direct-Net method proposed involves parameters and their value selections.

7d. The recent one-step algorithms seek to accurately invert imaging model *equation (1)*, i.e., accurate image reconstruction, for various scanning configurations of potential interest in DECT and spectral CT. Therefore, the iteration number (i.e., computation time) is not a leading factor of research concern in terms of accurately inverting the imaging model. The recent one-step algorithms were compared with the standard IDD and DDD methods in studies with simulated noisy data and real data collected in experiments; and the comparison results demonstrate that the performance of the recent one-step algorithms is not inferior to that of the well-established standard IDD and DDD methods in full-scan DECT. Additionally, for partial-scan DECT considered, the recent one-step algorithms have also been demonstrated to invert accurately *equation (1)*, yielding image reconstruction and concentration estimation comparable to those obtained from data in full-scan DECT.\*\*\*

---------------------------------------------------------------------------------------------------------------------

**Footnotes:**

\* It is stated in the paper that ``*the degraded decomposition performance of DIRECT-JSI method is mainly caused by the inconsistencies between the assumed Poisson noise model in the algorithm and the real Gaussian-like noise distribution in the experiment*." The statement is unlikely to be correct, as it can be demonstrated in a straightforward study described below: Using noiseless data generated from the numerical phantoms with *equation (1)* in the paper for full-scan DECT, one can reconstruct images with the DIRECT-JSI and Direct-Net methods. Since there is no data noise in this deterministic case, the data-noise-statistics maximum-likelihood argument becomes irrelevant. If there remains a "*degraded decomposition performance of DIRECT-JSI method*" in the study, it cannot be attributed to the choice of noise models anymore; and it might be instead the result of other issues such as incorrect implementation of the methods. This study can be used indeed specifically for verifying the implementation

correctness of the methods because, in this noiseless, ideal case, they should yield, up to the computer precision, numerically identical, accurate reconstructions.

** The problem of image reconstruction from full-scan data in DECT is well-studied, just like the problem of image reconstruction from the full-scan Radon transform (RT) that is well-studied. Basic research on additional algorithms only for inverting the full-scan RT, i.e., only for image reconstruction from knowledge of the full-scan RT, seems of little relevance and significance.

*** The paper cites that "*the ID-EP method spent about 625 s for 25000 iterations*" and that "*the Direct-JSI method took about 627 s for 40 iterations.*"

First, the standard IDD method, which was not compared to, is by far the most widely used in clinical and other applications of DECT. The standard IDD method obtains basis images by directly inverting the 2X2 matrix of equation (1) in *Ref. [7]* [2] from the low- and high-kVp FBP reconstructions, certainly without any iterations. Indeed, the results in *Ref. [7]* [2] appear to show that the standard IDD and ID-EP methods are of a comparable level of performance in terms of image visualization and concentration estimation. Therefore, the statement "*the ID-EP method spent about 625 s for 25000 iterations*" can be misleading, as it may feed a false impression that any capable IDD methods would require intensive computation for basis image decomposition.

Second, the DIRECT-JSI method seeks to reconstruct images by solving the optimization problem, i.e., equation (8) in *Ref. [14]* [3]. As discussed in Comment 7d above, when its convergence is of concern in terms of solving an optimization problem, the number of iterations (and thus computation time) is not a leading factor of relevance.

Computation time of image reconstruction is clearly an extremely important consideration in a real-world application or product. However, it is a misconception to expect that theoretical algorithms reported in research articles in the literature can be applied directly to addressing optimally a real-world, practical reconstruction application, because the algorithms are not developed, optimized, and evaluated under conditions unique to the workflow and task-specific utility metrics of the real-world application. In fact, it is unlikely that the specific conditions and requirements in an application or product, and even the application (or product) itself, are known at the time an algorithm is investigated and developed as a research work published in the literature.

It is a reconstruction procedure, along with its specifically selected parameters, that is adopted in a practical application or product. The reconstruction procedure necessarily is tailored to, and optimized for, the unique workflow settings (such as fast reconstruction time,) subject and data conditions, and task-specific metric requirements in the application or product, even though the reconstruction procedure itself generally does not solve the image model that an algorithm is developed to solve in a research work published in the literature. It may be the case that the design of only some portions of a reconstruction procedure of practical utility is motivated by an algorithm published in the literature. To simply put, an algorithm published as a research work in the literature is generally unlikely to be equivalent to a reconstruction procedure optimized for DECT practical application or product. Evidence is not provided that the proposed Direct-Net method has been optimized and evaluated in any practical application or product.